\documentclass[final,5p,times,twocolumn]{elsarticle}

\usepackage{lineno,hyperref}

\usepackage{graphicx}
\usepackage{bm}
\usepackage{float}
 \usepackage{amsthm}
 \usepackage{amssymb}

\usepackage{graphicx,wrapfig,lipsum}
\usepackage{subeqnarray}
\usepackage{color}
\usepackage{eqnarray,amsmath}

\journal{Physics Letters B}
\begin{document} 

\begin{frontmatter}

%\preprint{arXiv:1810.06717 [hep-ph]}

\title{Cosmological Strangeness Abundance}

\author{Cheng Tao Yang}
\author{Johann Rafelski}
% \email{Second.Author@institution.edu}
\address{Department of Physics, The University of Arizona, Tucson, AZ 85721, USA}
\date{\today}

\begin{abstract}
We investigate the strange particle composition of the early Universe in the hadron epoch $T_h\approx 150\ge T\ge 10$\,MeV. We study strangeness yield in thermal and chemical equilibrium constrained by prescribed entropy per baryon in a charge neutral and strangeness neutral $\langle s-\bar s\rangle$ Universe. Turning to kinetic processes in a Hubble expanding Universe, we determine conditions at which individual strangeness producing reactions fall out of detailed balance between decay and back-reaction strangeness production rates in presence of decreasing temperature $T$; we allow for weak, electromagnetic, and strong interaction processes. The weak interaction $\mu^\pm+\nu_{\mu}\rightarrow K^\pm$ freezeout is at $T_f^{K^\pm}=33.8\,\mathrm{MeV}$; the electromagnetic process $l^-+l^+\rightarrow\phi$ freezeout is at $T_f^\phi=23\sim25\,\mathrm{MeV}$; and the hadronic reaction $\pi+\pi\rightarrow K$ freezeout is at $T_f^K=19.8\,\mathrm{MeV}$.
\end{abstract}
\end{frontmatter} 
%%%%%%%%%%%%%%%%%%%%%%%%%%%%%%%%%%%%%%%%%%%%%%%%%%%%%%%%%%%%%%%%%%
\section{Introduction}
In order to fully describe the nonequilibrium processes in the early Universe we explore here the strangeness abundance and decoupling after normal matter formation in hadronization of quark- gluon plasma (QGP). Nonequilibrium conditions in the early Universe are of general interest: they are understood to be prerequisite for the arrow of time dependent processes to take hold in the Hubble expanding Universe. The two well-studied cases are:\\
i) The Big Bang Nucleosynthesis~\cite{Pitrou:2018cgg,Kolb:1990vq,Dodelson:2003ft,Mukhanov:2005sc} (BBN) appears in the temperature range $ 0.07>T>0.01\,\mathrm{MeV}$, about 1000 times smaller compared to what will be considered here, $T_h\approx 150\ge T\ge 10$\,MeV. Certain light isotopes, {\it e.g.\/} Lithium, are residual small remnants of this Universe era; \\
ii) Baryogenesis is believed to occur at or before the Universe underwent electroweak (EW) phase transition~\cite{Kolb:1990vq} at a temperature $T\simeq 130$\, GeV, about 1000 times larger compared to the temperature range we consider. 

However, the expected EW baryogenesis rate is small and some additional mechanism is required: Heavy bottom flavor decoupling near to the deconfined QGP transformation to the hadron gas (HG) phase just at the upper range of our present consideration may offer alternate baryogenesis opportunity~\cite{Yang:2020nne}.

We begin presenting the equilibrium hadron content of the Universe~\cite{Fromerth:2012fe,Rafelski:2013yka}, allowing for the presence of strange quark flavor for the given present day entropy per baryon assumed to also govern the hadron era charge neutral and strangeness neural Universe. We use here standard methods developed in the study of QGP and HG properties created in relativistic heavy ion (RHI) collisions. Ref.\cite{Letessier:2002gp} provides appropriate introduction. 

Moving beyond equilibrium Universe, we generalize the RHI collision strangeness abundance study by Koch, M\"uller and Rafelski~\cite{Koch:1986ud} to include electromagnetic and weak interaction processes. We apply methods of kinetic theory seeking the temperature at which strangeness decouples and disappears from the Universe. We illustrate our approach considering an unstable strange particle, say $S$, decaying into two particles $S\rightarrow 1+2$. In a dense and high temperature plasma with particle $1$ and $2$ present in thermal equilibrium, the inverse reaction $1+2\rightarrow S$ produces the particle $S$. The natural decay of particles concerned provides also the intrinsic strength of the inverse, strangeness production reaction. 

As long as both decay and production reactions are possible, particle $S$ abundance remains in thermal equilibrium. This balance between production and decay rates is called detailed balance. We used this method before when considering the abundance (non)equilibrium of hadrons driven by reaction $\pi^0\Leftrightarrow \gamma+\gamma$~\cite{Kuznetsova:2008jt}, and we validate these early results in this work. 

Once the primordial Universe expansion rate, given as the inverse of the Hubble parameter $1/H$, overwhelms the strongly temperature dependent back-reaction, the decay $S\rightarrow 1+2$ occurs out of balance and particle $S$ disappears from the inventory. The two-on-two strangeness producing $S+3\leftrightarrow 1+2$ reactions have a significantly higher strangeness production reaction threshold, thus especially near to strangeness decoupling their influence is negligible. Such reactions are more important near the QGP hadronization temperature $T\simeq 150$\,MeV; and they characterize strangeness exchange reactions such as $\mathrm{K}+N\leftrightarrow \Lambda+\pi$, see Chapter 18 in Ref.\cite{Letessier:2002gp}, which we also explore.

%%%%%%%%%%%%%%%%%%%%%%%%%%%%%%%%%%%%%%%%%%%%%%%%%%%%%%%%%%%%%%%%%%

\section{Hubble expansion}

\begin{figure}[t]
%\begin{center}
\centering
\includegraphics[width=0.95\linewidth]{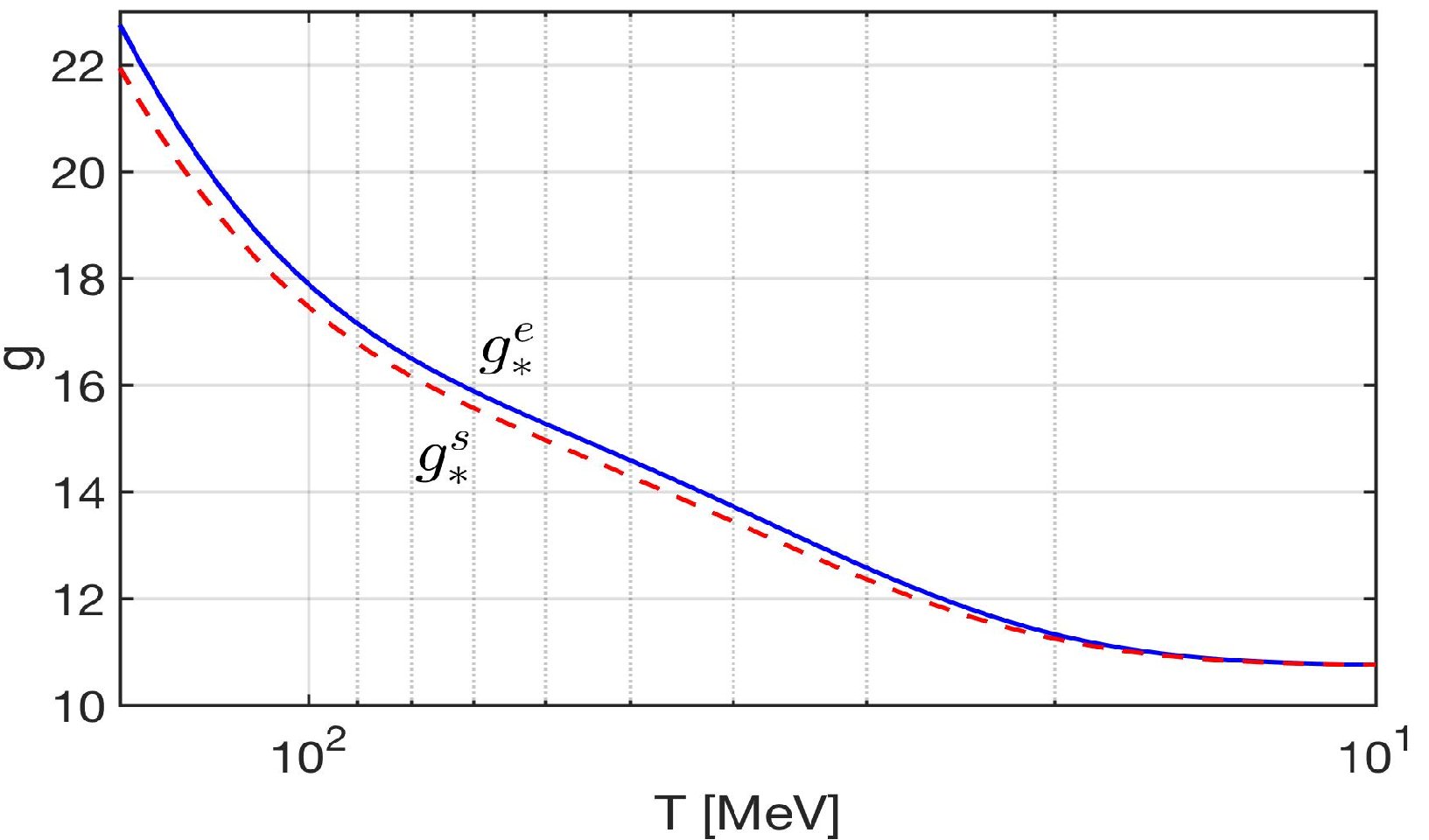}
\caption{For temperature $T$ in the early Universe epoch explored in this work $T_h\approx 150\ge T\ge 10$\,MeV we show the number of effective `energy' degrees of freedom $g^e_\ast$(solid, blue line), and number of effective 'entropy' degree of freedom $g^s_\ast$ (dashed, red line). }
\label{EntropyDOFFig} 
\end{figure}

An important assumption allowing us to explore the early Universe evolution is that both baryon and entropy content of the Universe is conserved in the comoving volume, scaling with the third power of the expansion parameter $a(t)$, where $H=\dot a/a$. We have not discovered any process capable of altering either entropy in the hadronic Universe appreciably. Baryon number is more strictly conserved. Therefore in this study we follow the usual assumption of an adiabatic Universe in which the ratio of baryon number density to visible matter entropy density remains constant during hadronic epoch
\begin{align}
\frac{n_B-n_{\overline{B}}}{\sigma}= \left.\frac{n_B-n_{\overline{B}}}{ \sigma}\right|_{t_0}=\mathrm{Const.}\;.
\end{align}
The subscript $t_0$ denotes the present day condition, assumed to be the same as in the hadronic era, and $\sigma$ is the total entropy density (photons, electrons, muons, neutrinos, pions and residual hadrons). In general the entropy density in the early Universe can be written as
\begin{align}
\sigma=\left(\frac{\rho+P}{T}-\sum_i\frac{\mu_i}{T}n_i\right)=\frac{2\pi^2}{45}g^s_\ast\,T^3,
\end{align}
where $g^s_\ast$ counts the effective number of `entropy' degrees of freedom. In Fig.\,\ref{EntropyDOFFig} $g^s_\ast$ is shown (dashed, red) as a function of temperature $T$. The effect of particle mass threshold~\cite{Coc:2006rt} is considered in the calculation for all involved particles. When $T$ decreases below the mass of particle $T\ll m_i$, this particle species becomes nonrelativistic; the contribution to $g^s_\ast$ becomes negligible, creating the dependence on $T$ seen in Fig.\,\ref{EntropyDOFFig}.

%~~~~~~~~~~~~~~~~~~~~~~~~~~~~~~~~~~~~~~~~~~~~~~~~~~~~~~~~~~~~~~~~~~~~~~~~~~~~~~~~
%~~~~~~~~~~~~~~~~~~~~~~~~~~~~~~~~~~~~~~~~~~~~~~~~~~~~~~~~~~~~~~~~~~~~~~~~~~~~~~

In the epoch of interest the Universe is dominated by radiation and effectively massless matter behaving like radiation. Therefore the Hubble parameter can be written as~\cite{Kolb:1990vq}
\begin{align}\label{H2g}
H^2=H^2_{rad}\left(1+\frac{\rho_{\pi,\,\mu,\,\rho}}{\rho_\mathrm{rad}}+\frac{\rho_\mathrm{strange}}{\rho_\mathrm{rad}}\right)=\frac{8\pi^3G_\mathrm{N}}{90}g^e_\ast T^4,
\end{align}
where: $g^e_\ast$ is the total number of effective relativistic `energy' degrees of freedom seen in Fig.\, \ref{EntropyDOFFig} (solid, blue) as a function of temperature $T$; $H^2_\mathrm{rad}=8\pi G_\mathrm{N}\,\rho_\mathrm{rad}/3$; $G_\mathrm{N}$ is the Newtonian constant of gravitation; the `radiation' energy density includes $\rho_\mathrm{rad}=\rho_\gamma+\rho_\nu+\rho_{e^\pm}= g^e\pi^2T^4/30$ with $g^e=10.75$ for photons, neutrinos, and massless electrons(positrons), see the limit of low $T$ in Fig.\, \ref{EntropyDOFFig}. We further note in Eq.\,(\ref{H2g}) a massive-particle correction $\rho_{\pi,\,\mu,\,\rho}=\rho_\pi+\rho_\mu+\rho_\rho$; and at highest $T$ of interest, also of (minor) relevance, $\rho_\mathrm{strange}=\rho_{K^0}+\rho_{K^\pm}+\rho_{K^\ast}+\rho_{\eta}+\rho_{\eta^\prime}$ is shown creating the behavior seen for high $T$ in Fig.\, \ref{EntropyDOFFig}.

In a kinetic model of neutrino decoupling no entropy generating processes were discovered~\cite{Birrell:2014gea,Birrell:2014uka}. For the hadronic Universe in the presence of detailed balance (dynamic equilibrium) between different reactions, no entropy can be produced. However, production of entropy can occur when the expansion of Universe breaks the detailed balance, {\it i.e.\/}, when the rate of production of a decaying particle is smaller compare to the rate of expansion of the Universe $H$. Once production reaction slows down and becomes ineffective in maintaining the dynamical equilibrium, we have the last decay for the unstable particle. 

On first sight particle number changing processes such as $K^0\rightarrow 1 +2 +\ldots $ imply that entropy could be now produced when particles fall out of dynamics equilibrium. However, decoupling occurs when the mass of the unstable particle is much larger than the ambient temperature. Therefore, the reactions involving a change of particle number occur when the `mother' particle abundance is very small. This implies that change in the entropy content of the Universe remains negligible. A more detailed study of entropy production will be presented in separate work~\cite{Yang:2021abc}.
 
In the epoch of interest and within the context of the $\Lambda$CDM standard model of the Universe, dark energy and massive dark matter contribute negligibly to the energy density: The cold dark matter (CDM) energy content emerges along with visible matter once radiation content decreased from once overwhelming contribution to energy inventory, allowing CDM to become the dominant matter component today. Dark energy is interpreted as the Einstein cosmological parameter $\Lambda$ emerging in the energy inventory very near to the present time epoch.

In Table\,\ref{TemperatureTime_table} we show for a sample of temperature values $T$ in the 1st column the characteristic Universe expansion time constant $H^{-1}$ in the 2nd column, the elapsed time $t_h$ in the Universe after hadronization in the 3rd column, and the number of e-folds $N$ in the third column. To obtain the elapsed time we have set the initial age of the Universe at onset of hadronization to be $45\,\mu\mathrm{s}$; by the time the Universe reaches the next sample point of $T=100$\,MeV our result is independent of this choice of initial condition. The Universe $N$-folding is defined as usual~\cite{ParticleDataGroup:2018ovx}
\begin{align}
N\equiv\ln\left(\frac{a(t_f)}{a(t_i)}\right)=\int_{t_i}^{t_f}H\,dt\;.
\end{align}
For a given value $N$ the Universe scale factor $a(t)$ has expanded by a factor $e^N$. The values of $N$ we see are near to the values expected for a radiation dominated Universe ($Ta=$Const.), in which case we would expect at $T=5$\,MeV a value $N=ln(150/5)=3.4$. The deviation $3.6>3.4$ arises during the epoch $5< T< 150$\,MeV since muons and pions present in very significant abundance at $T=150$\,MeV disappear from the inventory, reheating the Universe. 
 
We see in Table\,\ref{TemperatureTime_table} that the epoch $10< T< 150$\,MeV of interest to us lasts 7.4 milliseconds, which is in the realm of elementary processes is a much longer time interval than all natural time constants, except the lifespan of the neutron not of relevance in our study. This means that the particle decoupling from the primordial plasma arises because the back-particle formation of a decay process has slowed down at low temperature and is unable to keep up with the speed of Universe expansion.

%~~~~~~~~~~~~~~~~~~~~~~~~~~~~~~~~~~~~~~~~~~~~~~~~~~~~~~~~~~~~~~~~~~~~~~~~~~~~~~~~~~~~~~~~~\
\begin{table}%[h]
\caption{For a sample of values of temperature $T$ we show the characteristic Universe time constant $H^{-1}=\dot a/a$, the age of hadron phase $t_h$, and the number $N$ of e-folds.}
\label{TemperatureTime_table} 
\centering
\begin{tabular}{c| c| c| c}
\hline\hline
$T$\,[MeV] & $H^{-1}$\,[s] & $t_h$\,[s] &{N \,[e-folds]} \\
\hline
$150$ & $4.51\times10^{-5}$ & $0$ & {$0$}\\
\hline
$100$ & $1.14\times10^{-4}$ & $3.45\times10^{-5}$ &{$0.46$}\\
\hline
 $50$ & $5.07\times10^{-4}$ & $2.31\times10^{-4}$ & {$1.21$}\\
\hline
$10$ & $1.49\times10^{-2}$ & $7.44\times10^{-3}$ & {$2.90$}\\
\hline
$5$ & $5.90\times10^{-2}$ & $2.95\times10^{-2}$ & {$3.59$}\\
\hline\hline
\end{tabular}
\end{table}

%~~~~~~~~~~~~~~~~~~~~~~~~~~~~~~~~~~~~~~~~~~~~~~~~~~~~~~~~~~~~~~~~~~~~~~~~~ 

%%%%%%%%%%%%%%%%%%%%%%%%%%%%%%%%%%%%%%%%%%%%%%%%%%%%%%%%%%%%%%%%%%

\section{Chemical fugacities and particle abundances}
As long as back reactions are faster than the Universe expansion, which condition(s) we characterize in the following, we can explore the Universe composition assuming both kinetic and particle abundance equilibrium (chemical equilibrium). In prior works~\cite{Fromerth:2012fe,Rafelski:2013yka} charge neutrality and prescribed conserved baryon-per-entropy-ratio ${(n_B-n_{\overline{B}})}/{\sigma}$ were used to determine the baryochemical potential $\mu_B$. 

The net baryon density in early Universe with temperature range $10< T< 150$\,MeV can be written as
\begin{align}
\frac{\left(n_B-n_{\overline{B}}\right)}{\sigma}&=\frac{1}{\sigma}\left[\left(n_p-n_{\overline{p}}\right)+\left(n_n-n_{\overline{n}}\right)+\left(n_Y-n_{\overline{Y}}\right)\right]\notag\\&=\frac{45}{4\pi^4g^s_\ast}\left(\lambda_q^3-\lambda_q^{-3}\right)\left[F_N-\frac{\lambda_s}{\lambda_q}\,F_Y\right],
\end{align}
where the variables $\lambda_s$ and $\lambda_q$ are related to chemical potential of strangeness and quark $\lambda_s=\exp(\mu_s/T)$, and $\lambda_q=\exp(\mu_B/3T)$. The phase-space function $F_i$ for sets of nucleon $N$, kaon $K$, and hyperon $Y$ particles is 
\begin{align}
&F_i=\sum_i\,g_{i}\left(\frac{m_{i}}{T}\right)^2\,K_2(m_{i}/T)\;, \quad i=N,K,Y\;.
\end{align}
We include nucleons $N_i=n, p, \Delta(1232)$, hyperons $Y_i=\Lambda, \Sigma^0,\Sigma^\pm, \Sigma(1385)$, and kaons $K_i=K^0, \overline{K^0}, K^\pm, K^\ast(892)$. Imposing $\langle s-\bar s \rangle=0$, when the baryon chemical potential does not vanish the chemical potential of strangeness in the early Universe, satisfies (see Section 11.5 in Ref.\,\cite{Letessier:2002gp})
\begin{align}\label{museq}
\lambda_s=\lambda_q\sqrt{\frac{F_K+\lambda^{-3}_q\,F_Y}{F_K+\lambda^3_q\,F_Y}}.
\end{align}
 
Introducing  strangeness $\langle s-\bar s\rangle=0$ constraint, the explicit relation for baryon to entropy ratio is
\begin{align}\label{muBeq}
\frac{n_B-n_{\overline{B}}}{\sigma}&=\frac{45}{2\pi^4g^s_\ast}\sinh\left[\frac{\mu_B}{T}\right]F_N\notag\\
\times &\left[1-\frac{F_Y}{F_N}\sqrt{\frac{1+e^{-\mu_B/T}\,F_Y/F_K}{1+e^{\mu_B/T}\,F_Y/F_K}}\right].
\end{align}
Governing Eq.\,(\ref{muBeq}) is the present-day baryon-per-entropy-ratio, and we obtain the value 
\begin{align}\label{BdS}
\frac{n_B-n_{\overline{B}}}{\sigma}= \left.\frac{n_B-n_{\overline{B}}}{ \sigma}\right|_{t_0}=(0.865\pm0.008)\times10^{-10} \;.
\end{align}
The value we obtained in Eq.\,(\ref{BdS}) is 12\% larger compared to the year 2003 Ref.\,\cite{Fromerth:2012fe} where $0.77\pm 0.06\times10^{-10}$ was used; this difference is due to a change in input parameters describing properties of the Universe. For a detailed evaluation method we refer to this earlier work now using a baryon-to-photon ratio~\cite{ParticleDataGroup:2018ovx}: $\left(n_B-n_{\overline{B}}\right)/n_\gamma= (0.609\pm0.06)\times10^{-9}$, as well as the entropy per particle for a massless boson $\sigma/n|_\mathrm{boson}\approx 3.60$ and massless fermion $\sigma/n|_\mathrm{fermion}\approx 4.20$. 

Neutrinos, a free-streaming cosmic plasma component, are out of the thermal equilibrium and the conventional relation between entropy, energy density, and pressure does not apply. Since neutrino decoupling occurs at $T_k\gg m_\nu$, the entropy content of free-streaming neutrinos is computed as if neutrinos were massless~\cite{Birrell:2012gg}, even when $T_\gamma<m_\nu$. In the evaluation of baryon-per-entropy-ratio we allow for the reheating of photons by $e^+e^-$ annihilation, thus the entropy content of photons incorporates that of the $e^+e^-$ primordial plasma. 

%~~~~~~~~~~~~~~~~~~~~~~~~~~~~~~~~~~~~~~~~~~~~~~~~~~~~~~~~~~~~~~~~~~~~~~~~~~~~~~~~
\begin{figure}[t]
%\begin{center}
\centering
\includegraphics[width=0.95\linewidth]{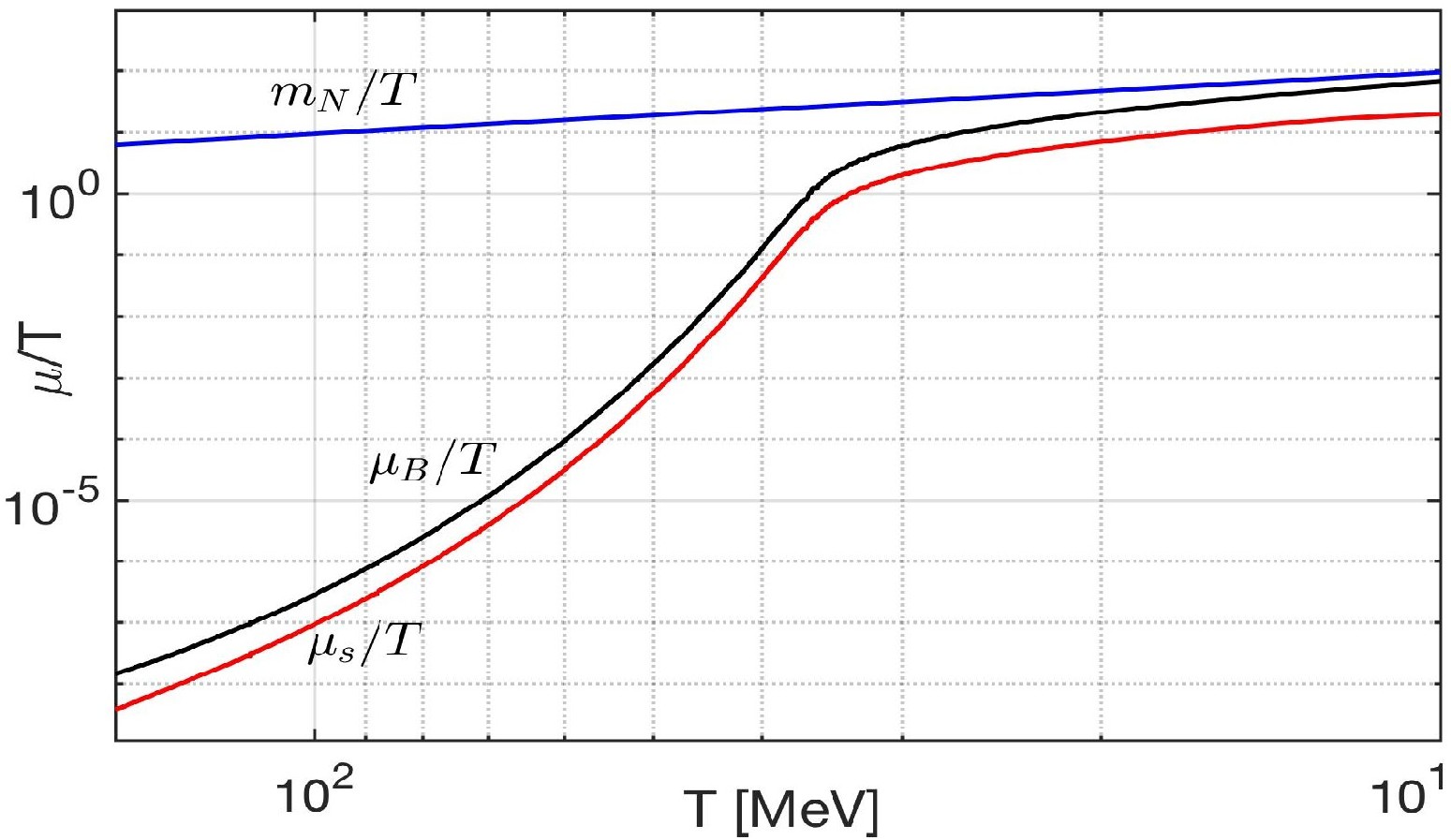}
\caption{The chemical potential of baryon $\mu_B/T$ and strangeness $\mu_s/T$ as a function of temperature $150\,\mathrm{MeV}> T>10\,\mathrm{MeV}$ in the early Universe; for comparison we show $m_N/T $ with $m_N=938.92$\,MeV, the average nucleon mass.}
\label{ChemPotFig}
%\end{center}
\end{figure}
\begin{figure}[bt]
%\begin{center}
\centering
\includegraphics[width=0.95\linewidth]{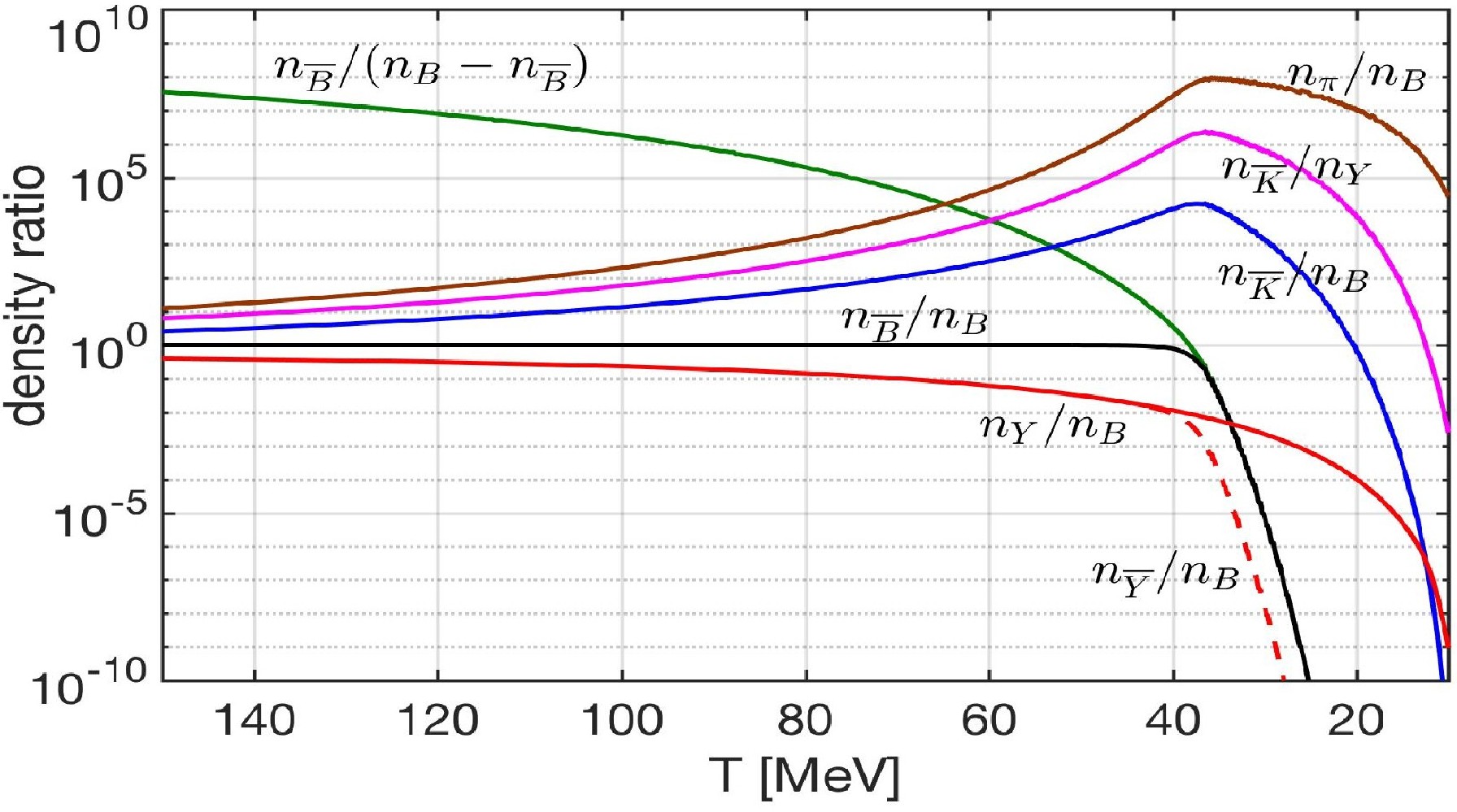}
\caption{Ratios of hadronic particle number densities as a function of temperature $150\,\mathrm{MeV}> T>10\,\mathrm{MeV}$ in the early Universe with baryon $B$ yields: pions $\pi$ (brown line), kaons $K( q\bar s)$ (blue), antibaryon $\overline B$ (black), hyperon $Y$ (red) and anti-hyperons $\overline Y$ (dashed red). Also shown $\overline K/Y$(purple).}
\label{EquilibPartRatiosFig}
%\end{center}
\end{figure}

%~~~~~~~~~~~~~~~~~~~~~~~~~~~~~~~~~~~~~~~~~~~~~~~~~~~~~~~~~~~~~~~~~~~~~~~~~~~~~~
We solve Eqs.\,(\,\ref{museq}),\,(\ref{muBeq})  numerically to obtain Fig.~\ref{ChemPotFig} as a function of $T$. Our results agree with Ref.\,\cite{Fromerth:2012fe}, allowing for a slightly different set of cosmological parameters, as noted above. The chemical potentials changes dramatically in the temperature window $50\le T\le 30$\,MeV, which behavior is describing the process of antibaryon disappearance. In Fig.~\ref{EquilibPartRatiosFig} we show examples of particle abundance ratios. Of special interest is the temperature where antibaryons disappear from the Universe inventory, defined when the ratio $n_{\overline B}/(n_B-n_{\overline B})=1$. This condition is reached in an expanding Universe at $T=38.2$\,MeV. Considering $n_Y/n_B$ we see that hyperons $Y(sqq)$ remain a noticeable 1\% component in baryon yield through this domain of antibaryon decoupling.

%~~~~~~~~~~~~~~~~~~~~~~~~~~~~~~~~~~~~~~~~~~~~~~~~~~~~~~~~~~~~~~~~~~~~~~~~~~~~~~~~
%~~~~~~~~~~~~~~~~~~~~~~~~~~~~~~~~~~~~~~~~~~~~~~~~~~~~~~~~~~~~~~~~~~~~~~~~~~~~~~

Pions $\pi(q\bar q)$ are the most abundant hadrons because of their low mass and the reaction $\gamma\gamma\rightarrow\pi^0$, which assures chemical yield equilibrium~\cite{Kuznetsova:2008jt}; see brown line for $n_\pi/n_B\gg1$ in Fig.~\ref{EquilibPartRatiosFig}. We can understand some of counterintuitive results appearing at low temperatures by remembering that the threshold energy for strangeness in $\Lambda(uds)$ costs less energy compared to $K( q\bar s)$: For $150\,\mathrm{MeV}>T>20\,\mathrm{MeV}$ we see the ratio $n_{{\overline K}(\bar q s)}/n_B\gg1$ which implies pair abundance of strangeness is more abundant than baryons, and is dominantly present in mesons, since $n_{\overline K}/n_Y\gg1$; for $12.9\,\mathrm{MeV}>T$ we have $n_Y/n_B>n_{\overline K}/n_B$, now the still existent tiny abundance of strangeness is found predominantly in hyperons. We find below that the exchange reaction $\overline{K}+N\rightarrow \Lambda+\pi$ can re-equilibrate kaons and hyperons in the temperature range; therefore strangeness symmetry $s=\bar s$ is maintained.

%%%%%%%%%%%%%%%%%%%%%%%%%%%%%%%%%%%%%%%%%%%%%%%%%%%%%%%%%%%%%%%%%%

\section{Strangness abundance decoupling} 
The reaction rates for inelastic collision process capable of changing particle number, for example $\pi\pi\to K^0$, is suppressed by the factor $\exp{(-m_{K^0}/T)}$. On the other hand, there is no suppression for the elastic momentum and energy exchanging particle collisions in plasma. We conclude that for the case $m\gg T$, the dominant collision term in the relativistic Boltzmann equation is the elastic collision term keeping all heavy particles in kinetic energy  equilibrium with the plasma. This allows to study the particle abundance in plasma presuming the energy-momentum statistical distribution equilibrium exists. This insight was in detail discussed in the preparator phase of laboratory exploration of hot hadron and quark matter, see Ref.\,\cite{Koch:1986ud}.

In order to study the particle abundance in the Universe when $m\gg T$, instead of solving the exact Boltzmann equation ,we can separate the fast energy-momentum equilibrating collisions from the slow particle number changing inelastic collisions. In the following we explore alone the rates of inelastic collision and compare the relaxation times of particle production in all relevant reactions, one with another and with the Universe expansion rate.

\begin{figure} %[h]
%\begin{center}
\centering
\includegraphics[width=0.95\linewidth]{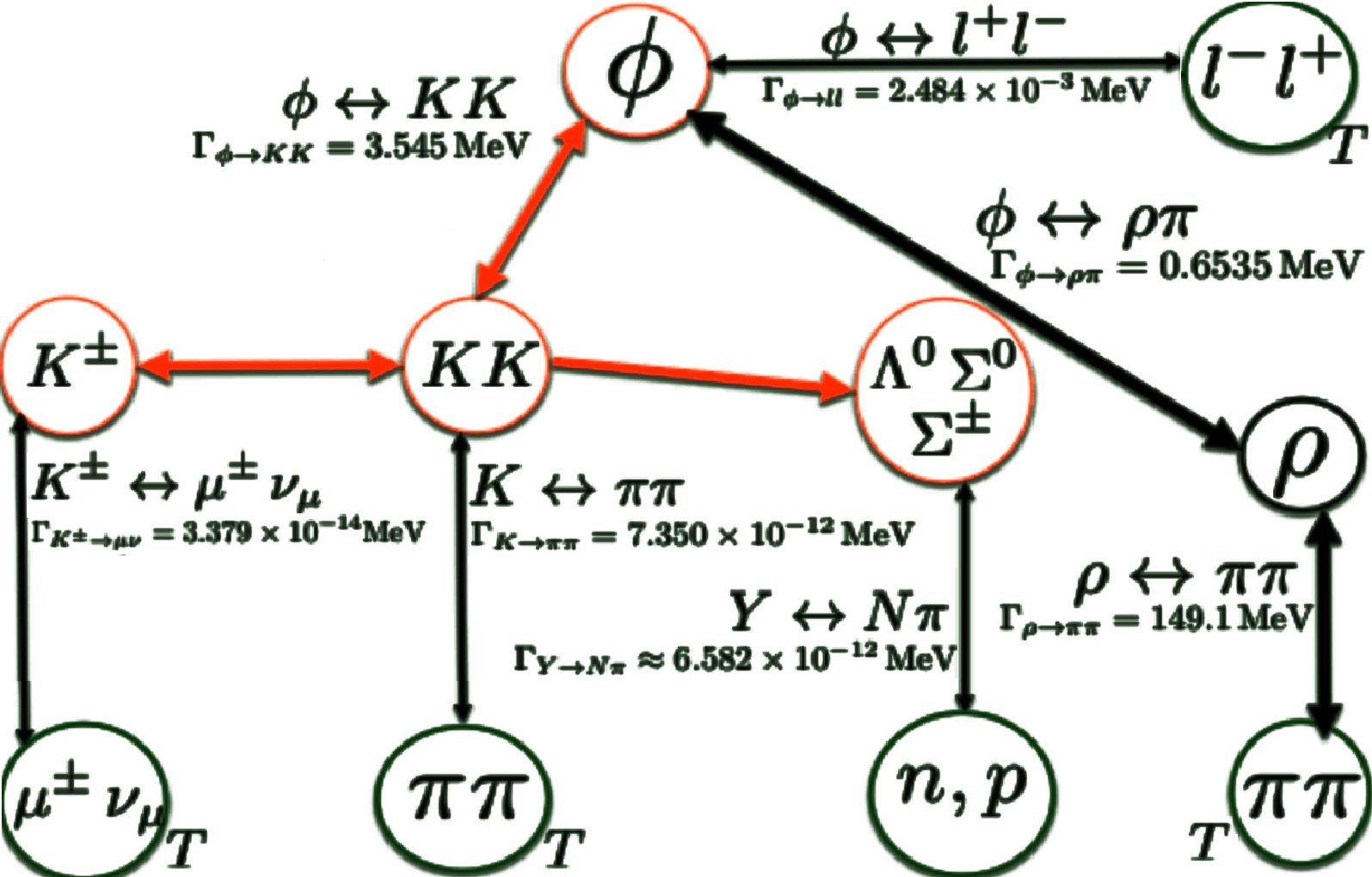}
\caption{The strangeness abundance changing reactions in the primordial Universe. The red circles show strangeness carrying hadronic particles; red thick lines denote effectively instantaneous reactions. Black thick lines show relatively strong hadronic reactions.}
\label{Strangeness_map2}
%\end{center}
\end{figure}

In order to determine where exactly strangeness disappears from the Universe inventory we explore the magnitudes of a relatively large number of different rates of production and decay processes, and compare these with the Hubble time constant.
The reactions at some point relevant to strangeness evolution in the considered Universe evolution epoch $T_h\approx 150\ge T\ge 10$\,MeV are illustrated in Fig.~\ref{Strangeness_map2} and the pertinent reaction strength is indicated. As shown:
\begin{itemize}
\item
We study strange quark abundance in baryons and mesons, considering both open and hidden strangeness (hidden: $s\bar s$-content). Important source reactions are $l^-+l^+\rightarrow\phi$, $\rho+\pi\rightarrow\phi$, $\pi+\pi\rightarrow K_\mathrm{S}$, $\Lambda \leftrightarrow \pi+ N$, and $\mu^\pm+\nu\rightarrow K^\pm$. 
\item
Muons and pions are coupled through electromagnetic reactions $\mu^++\mu^-\leftrightarrow\gamma+\gamma$ and $\pi\leftrightarrow\gamma+\gamma$ to the photon background and retain their chemical equilibrium until the temperature $T =4$\, MeV and $T=5$\,MeV, respectively~\cite{Rafelski:2021aey,Kuznetsova:2008jt}. The large $\phi\leftrightarrow K+K$ rate assures $\phi$ and $K$ are in relative chemical equilibrium.
\end{itemize}

%~~~~~~~Figure~~~~~~~~~~~~~~~~~~~~~~~~~~~~~~~~~~~~~~~~~~~~~~~~~~~~~~~~~~~~~~~~~~~~
%~~~~~~~~~~~~~~~~~~~~~~~~~~~~~~~~~~~~~~~~~~~~~~~~~~~~~~~~~~~~~~~~~~~~~~~~~~
%%%%%%%%%%%%%%%%%%%%%%%%%%%%%%%%%%%%%%%%%%%%%%%%%%%%%%%%%%%%%%%%%%%\

\subsection{Strangness in meson era}
The thermal reaction rate per time and volume for two body-to-one particle reactions $1+2\rightarrow 3$ has been presented before~\cite{Koch:1986ud,Kuznetsova:2008jt,Kuznetsova:2010pi}. In full kinetic and chemical equilibrium, the reaction rate per time per volume is given by:\cite{Kuznetsova:2010pi}
\begin{align}
&R_{12\to 3}=\frac{g_3}{(2\pi)^2}\,\frac{m_3}{\tau^0_3}\,\int^\infty_0\frac{p^2_3dp_3}{E_3}\frac{e^{E_3/T}}{e^{E_3/T}\pm1}\Phi(p_3)\;,
\end{align}
where $\tau^0_3$ is the vacuum lifetime of particle $3$. The positive sign $``+"$ is for the case when particle $3$ is a boson, and negative sign $``-"$ for fermion. The function $\Phi(p_3)$ for the non-relativistic limit $m_3\gg p_3,T$ can be written as 
\begin{align}
\Phi(p_3\to0)=2\frac{1}{(e^{E_1/T}\pm1)(e^{E_2/T}\pm1)}.
\end{align}

Considering the Boltzmann limit, the thermal reaction rate per unit time and volume becomes
\begin{align}
\label{Thermal_Rate}
R_{12\rightarrow3}=\frac{g_3}{2\pi^2}\left(\frac{T^3}{\tau^0_3}\right)\left(\frac{m_3}{T}\right)^2\,K_1(m_3/T).
\end{align}
In order to compare the reaction time with Hubble time $1/H$, it is convenient to define the relaxation time for the process $1+2\rightarrow 3$ as follows
\begin{align}
\label{Reaction_Time}
\tau_{12\rightarrow 3}\equiv\frac{n^{eq}_{1}}{R_{12\rightarrow n}},\;\;n^{eq}_1=\frac{g_1}{2\pi^2}\int_{m_1}^\infty\,dE\,\frac{E\,\sqrt{E^2-m_1^2}}{\exp{\left(E/T\right)}\pm1}\;, 
\end{align}
where $n^{eq}_1$\,is the thermal equilibrium number density of particle\,$1$.

It is common to refer to particle freeze-out as the epoch where a given type of particle ceases to interact with other particles. In an expanding Universe this can happen when the density of particles becomes too low because of the Universe expansion and the probability for two-particles collision is negligible, or when the particle energy becomes too small to overcome reaction threshold. In this situation the particle yield decouples from the cosmic plasma, a chemical nonequilibrium and even complete abundance disappearance of this particle follow; the condition for the given reaction $1+2\rightarrow 3$ to decouple is
\begin{align}
\tau_{12\rightarrow 3}(T_f)=1/H(T_f).
\end{align}

\begin{figure}[ht]
%\begin{center}
\centering
\includegraphics[width=0.95\linewidth]{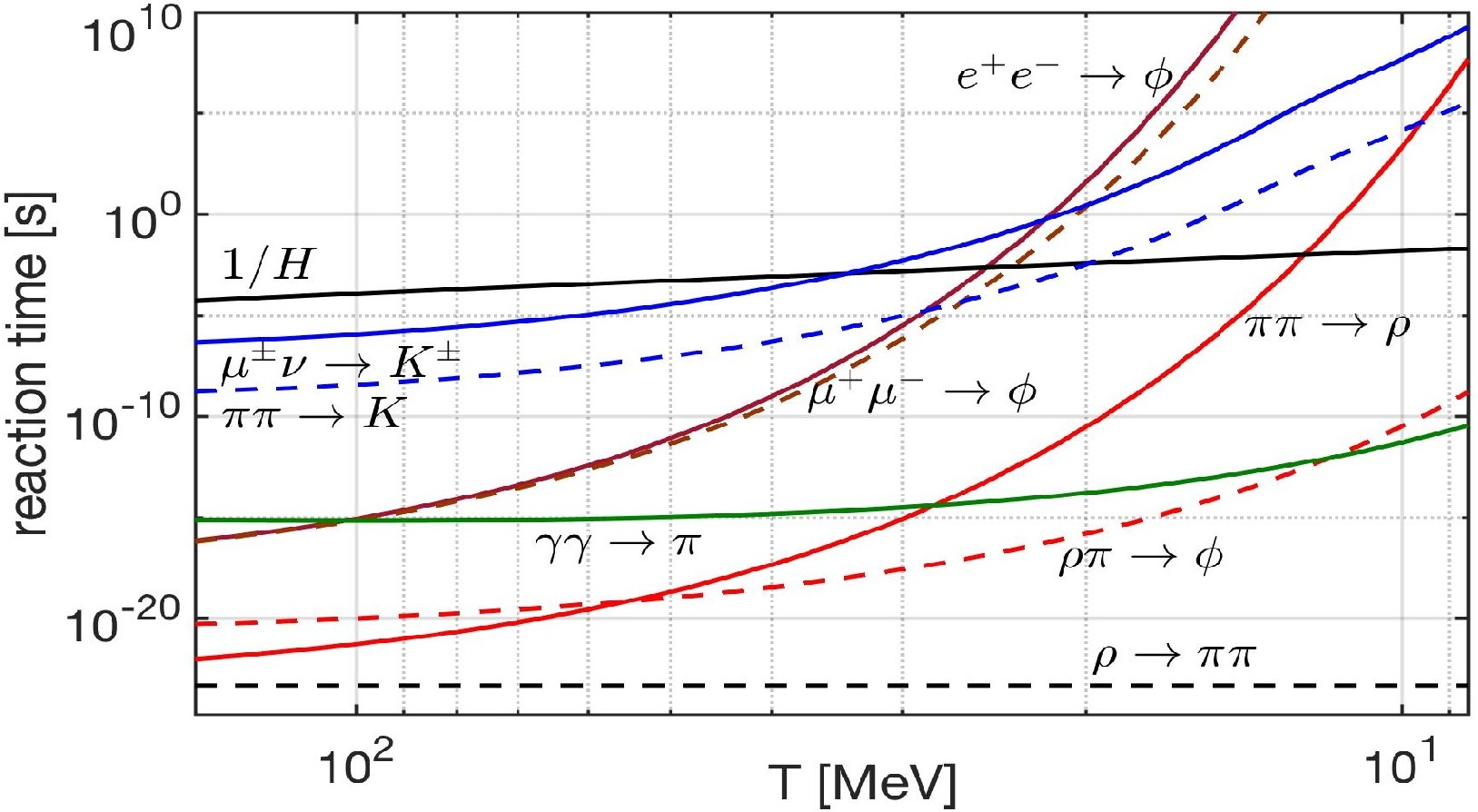}
\caption{Hadronic relaxation reaction times, see Eq.\,(\ref{Reaction_Time}), as a function of temperature $T$, are compared to Hubble time $1/H$ (black solid line). At bottom the horizontal black-dashed line is the natural (vacuum) lifespan of $\rho$.}
\label{reaction_time_tot}
%\end{center}
\end{figure}

When presenting the reaction rates and quoting decoupling as a function of temperature $T$ we must remember that for a temperature range $50>T>5$\,MeV, we have $10^{-1}<dT/dt<10^{-4}$\,MeV/$\mu$s. Therefore when hadronic rates cross the Hubble expansion rate in Fig.~\ref{reaction_time_tot} at a specific $T$, the ensuing particle disappearance occurs on a time scale so fast that in a $T$-dependence diagram this appears with graphic resolution comparable to line width. Therefore reading off the corresponding value of $T$ is sufficient to describe the domain of decoupling.

The relevant interaction rates competing with Hubble time involving strongly interacting mesons are the reactions $\pi+\pi\leftrightarrow K$, $\mu^\pm+\nu\rightarrow K^\pm$, $l^++l^-\rightarrow\phi$, $\rho+\pi\leftrightarrow\phi$, and $\pi+\pi\leftrightarrow\rho$. These rates are compared with Hubble time in Fig.~\ref{reaction_time_tot}.%~~~~~~~Figure~~~~~~~~~~~~~~~~~~~~~~~~~~~~~~~~~~~~~~~~~~~~~~~~~~~~~~~~~~~~~~~~~~~~~~~~~~
%~~~~~~~~~~~~~~~~~~~~~~~~~~~~~~~~~~~~~~~~~~~~~~~~~~~~~~~~~~~~~~~~~~~~~~~~~~~~~~~~~~~~~

 We see that the weak interaction (WI) reaction $\mu^\pm+\nu_{\mu}\rightarrow K^\pm$ becomes slower compared to the Universe expansion near temperature $T_f^{K^\pm}=33.8\,\mathrm{MeV}$ signaling the onset of abundance nonequilibrium for $K^\pm$. For $T<T_f^{K^\pm}$, the reactions $\mu^\pm+\nu_{\mu}\rightarrow K^\pm$ decouples from the cosmic plasma; the corresponding detailed balance can be broken and the decay reactions $K^\pm\rightarrow\mu^\pm+\nu_{\mu}$ are acting like a (small) ``hole'' in the strangeness abundance ``pot''. If other strangeness production reactions did not exist, strangeness would disappear as the Universe cools below $T_f^{K^\pm}$. However, we have other reactions: $l^++l^-\leftrightarrow\phi$, $\pi+\pi\leftrightarrow K$, and $\rho+\pi\leftrightarrow\phi$ can still produce the strangeness in cosmic plasma and the rate is very large compared to the WI small decay hole.

In Table\,\ref{FreezeoutTemperature_table} we show the characteristic strangeness reaction and their freezeout temperatures in the early Universe. The intersection of strangeness reaction times with $1/H$ occurs for $l^-+l^+\rightarrow\phi$ at $T_f^\phi=25\sim23\,\mathrm{MeV}$, and for $\pi+\pi\rightarrow K$ at $T_f^K=19.8\,\mathrm{MeV}$, for $\pi+\pi\rightarrow\rho$ at $T_f^\rho=12.3\,\mathrm{MeV}$. The reactions $\gamma+\gamma\rightarrow\pi$ and $\rho+\pi\leftrightarrow\phi$ are faster compared to $1/H$. However, the $\rho\to\pi+\pi$ lifetime (black dashed line in Fig.~\ref{reaction_time_tot}) is smaller than the reaction $\rho+\pi\leftrightarrow\phi$; in this case, most of $\rho$-meson decays faster, thus are absent and cannot contribute to the strangeness creation in the meson sector. Below the temperature $T<20$\,MeV, all the detail balances in the strange meson reactions are broken and the strangeness in the meson sector should disappear rapidly, were it not for the small number of baryons present in the Universe.

%~~~~~~~~~~~~~~~~~~~~~~~~~~~~~~~~~~~~~~~~~~~~~~~~~~~~~~~~~~~~~~~~~~~~~~~~~~~~~~~~~~~~~~~~~
\begin{table}%[h]
\caption{The characteristic strangeness reaction and their freezeout temperature in early Universe.}
\label{FreezeoutTemperature_table} 
\centering
\begin{tabular}{c| c}
\hline\hline
Reactions &Freezeout Temperature (MeV) \\
\hline
$\mu^\pm\nu\rightarrow K^\pm$ & $T_f=33.8$\,MeV\\
\hline
$e^+e^-\rightarrow \phi$ & $T_f=24.9$\,MeV\\
$\mu^+\mu^-\rightarrow\phi$ & $T_f=23.5$\,MeV\\
\hline
 $\pi\pi\rightarrow K$ & $T_f=19.8$\,MeV\\
\hline
$\pi\pi\rightarrow\rho$ & $T_f=12.3$\,MeV\\
\hline\hline
\end{tabular}
\end{table}

%~~~~~~~~~~~~~~~~~~~~~~~~~~~~~~~~~~~~~~~~~~~~~~~~~~~~~~~~~~~~~~~~~~~~~~~~~ 

%%%%%%%%%%%%%%%%%%%%%%%%%%%%%%%%%%%%%%%%%%%%%%%%%%%%%%%%%%%%%%%%%%%
 
\subsection{Strangeness in hyperons}
 We now consider the strangeness production reaction $\pi +N\rightarrow K+\Lambda$, the strangeness exchange reaction $\overline{K}+N\rightarrow \Lambda+\pi$; and the strangeness decay $\Lambda\rightarrow N+\pi$ allowing strange hyperons and anti hyperons to influence the dynamic nonequilibrium condition including development of $\langle s-\bar s\rangle \ne 0$. The cross sections $\sigma_{\overline{K}N\rightarrow \Lambda\pi}$ and $\sigma_{\pi N\rightarrow K\Lambda}$ are obtained from experiment. The thermal averaged cross sections for the strangeness production and exchange processes are about $\sigma_{\pi N\rightarrow K\Lambda}\sim0.1\,\mathrm{mb}$ and $\sigma_{\overline{K}N\rightarrow \Lambda\pi}=1\sim3\,\mathrm{mb}$ in the energy range we are interested. In a more detailed study we adopt the parameterization from papers~\cite{Koch:1986ud,Cugnon:1984pm} in our calculation.

\begin{figure}[h]
%\begin{center}
\centering
\includegraphics[width=0.95\linewidth]{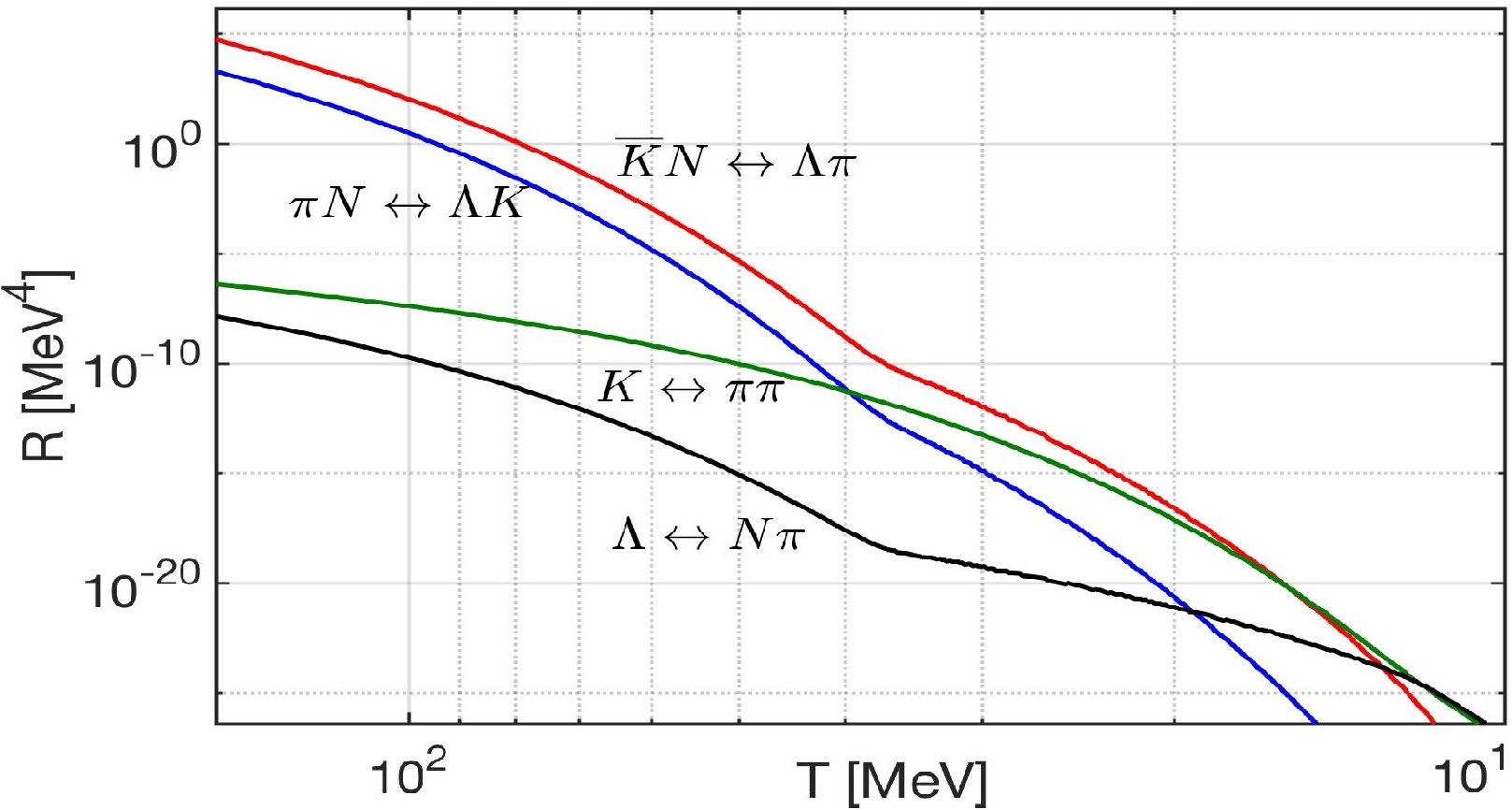}
\caption{Thermal reaction rate $R$ per volume and time for important hadronic strangeness production and exchange processes as a function of temperature $150\,\mathrm{MeV}> T>10\,\mathrm{MeV}$ in the early Universe.}
\label{Lambda_Rate_volume.fig}
%\end{center}
\end{figure}

The general form for thermal reaction rate per volume is discussed in~\cite{Letessier:2002gp} (Eq.(17.16), Chapter 17). Given the cross sections, we obtain the thermal reaction rate per volume for strangeness exchange reaction seen in Fig.~\ref{Lambda_Rate_volume.fig}. We see that around $T=20$\,MeV, the dominant reactions for the hyperon $\Lambda$ production is $\overline{K}+N\leftrightarrow\Lambda+\pi$. At the same time, the $\pi+\pi\to K$ reaction becomes slower than Hubble time and kaon $K$ decay rapidly in the early Universe. However, the anti-kaons $\overline K$ become the hyperon $\Lambda$ because the strangeness exchange reaction $\overline{K}+N\rightarrow\Lambda+\pi$ in the baryon-dominated Universe. We have strangeness in $\Lambda$ and it disappears from the Universe via the decay $\Lambda\rightarrow N+\pi$. Both strangeness and anti-strangeness disappear because of the $K\rightarrow\pi+\pi$ and $\Lambda\rightarrow N+\pi$, while the strangeness abundance $s = \bar{s}$ in the early Universe remains.

%~~~~~~~~~~~~~~~~~~~~~~~~~~~~~~~~~~~~~~~~~~~~~~~~~~~~~~~~~~~~~~~~~~~~~~~~~~~~~~~~
%~~~~~~~~~~~~~~~~~~~~~~~~~~~~~~~~~~~~~~~~~~~~~~~~~~~~~~~~~~~~~~~~~~~~~~~~~~~~~~

Around $T=12.9$\,MeV the reaction $\Lambda+\pi\rightarrow\overline{K}+N$ becomes slower than the strangeness decay $\Lambda\leftrightarrow N+\pi$, which shows that at the low temperature the $\Lambda$ particles are still in equilibrium via the reaction $\Lambda\leftrightarrow N+\pi$ and little strangeness remains in the $\Lambda$. Then strangeness abundance becomes asymmetric $s\gg \bar{s}$, which shows that the assumption for strangeness conservation can only be valid until the temperature $T\sim13$\,MeV. Below this temperature a new regime opens up in which the tiny residual strangness abundance is governed by weak decays with no re-equilibration with mesons. Also, in view of baron asymmetry, $\langle s-\bar s\rangle \ne 0$.

%%%%%%%%%%%%%%%%%%%%%%%%%%%%%%%%%%%%%%%%%%%%%%%%%%%%%%%%%%%%%%%%%%%
\section{Discussion}
We have presented results characterizing the evolution of cosmological strangeness abundance. Our work refines the understanding of the physical phenomena connecting the QGP phase with the neutrino decoupling stage~\cite{Fromerth:2012fe,Rafelski:2013yka}. The cosmic neutrino abundance~\cite{Birrell:2014uka} and free-streaming neutrino momentum distribution~\cite{Birrell:2012gg} could influence the speed of Universe expansion~\cite{Birrell:2014cja}, a topic generating a lot of interest today~\cite{Riess:2020sih,Verde:2019ivm}. Our work contributes insights about the connection between the QGP era and the neutrino decoupling era.

In the QGP epoch before hadronization at temperature $T>150$\,MeV, the strangeness formation processes explored in the laboratory~\cite{Letessier:2002gp} are fast enough to assure chemical equilibrium. In the phase transformation to hadrons the excess QGP entropy as compared to HG is absorbed in additional Universe comoving volume expansion, while excess strangeness has time to reequilibrate into equilibrium HG abundance, the governing time scales were introduced in Table~\ref{TemperatureTime_table}. This situation is opposite to what one encounters in the fast and explosive disintegration of QGP drop created in RHI collisions in the laboratory. The Universe is undergoing a relatively slow on hadronic time scale phase transformation emerging from QGP near to chemical equilibrium abundance of all particles here considered. Tracking the following particle abundance we adopted ratio of baryon to entropy content of the Universe surrounding us today. 

Before turning to non-equilibrium processes we have evaluated the chemical equilibrium composition in the following Universe expansion and found several Universe epochs of interest:
\begin{itemize}
\item
In the temperature range $20 <T <60$\,MeV the Universe is rich in physics phenomena involving strange mesons, (anti)baryons including (anti)hyperon abundances. Considering the inventory of the Universe strange mesons and baryons, we have reevaluated the temperature of the baryon disappearance, which is estimated qualitatively in Ref.\,\cite{Kolb:1990vq} to be $T=40$-$50\,\mathrm{MeV}$. We have seen in Fig.~\ref{EquilibPartRatiosFig} (black line) that antibaryons (sum of non-strange and strange anti-hyperons) effectively disappear just below $T=40$\,MeV. 
\item
We have established in Fig.~\ref{EquilibPartRatiosFig} the regimes in which strangness is predominantly in mesons or in baryons. We found that for temperature $150 \mathrm{MeV}>T>20\,\mathrm{MeV}$ the Universe is meson-dominant and the strangeness is dominantly present in the meson sector with $s=\bar s$. For temperature $T<20$\,MeV, the Universe becomes baryon-dominant. Below temperature $T<13$\,MeV, strangeness is present dominantly in hyperons, hence $s -\bar s\ne 0$.
\end{itemize}

We have evaluated reaction rates involving (strange) mesons and baryons in detail and found Universe epochs of interest that follow the strength of weak, electromagnetic and strong interactions:
\begin{itemize}
\item
The first reaction to become slower compared to Hubble time $1/H$ is the weak interaction $\mu^\pm+\nu_{\mu}\rightarrow K^\pm$ at $T_f^{K^\pm}=33.8\,\mathrm{MeV}$.
\item 
This is followed by the electromagnetic process $l^-+l^+\rightarrow\phi$ at $T_f^\phi=23\sim25\,\mathrm{MeV}$.
\item
At $T_f^K=19.8\,\mathrm{MeV}$ the hadronic reaction $\pi+\pi\rightarrow K$ becomes slower than the Hubble expansion. In this case, the strangeness in mesons would disappear for $T<20$MeV. 
\end{itemize}

In order to understand strangeness in hyperons, we evaluated the reaction $\pi +N\rightarrow K+\Lambda$, the strangeness exchange reaction $\overline{K}+N\rightarrow \Lambda+\pi$, and the strangeness decay $\Lambda\rightarrow N+\pi$, in detail. In Fig.(\ref{Lambda_Rate_volume.fig}) we saw that for $T<20$\,MeV, the reactions for the hyperon $\Lambda$ production is dominated by $\overline{K}+N\leftrightarrow\Lambda+\pi$. Both strangeness and anti-strangeness disappear from the Universe via the reactions $\Lambda\rightarrow N+\pi$ and $K\to\pi+\pi$, keeping the $s=\bar s$. Beginning with $T=12.9$ MeV, the dominant reaction is $\Lambda\leftrightarrow N+\pi$, which shows that at a lower temperature we still have (very little) strangeness remnant in the $\Lambda$. In this case, the strangeness abundance becomes asymmetric and we have $s\gg\bar{s}$ in the early Universe. Hence, strange hyperons and anti hyperons could enter into dynamic nonequilibrium condition including $\langle s-\bar s\rangle \ne 0$.

Our results depend on the baryon-per-entropy ratio ${(n_B-n_{\overline{B}})}/{\sigma}$ assumed to be constant during Universe evolution. The constant baryon-per-entropy-ratio is equivalent to the statement that the Universe evolves adiabatically, {\it i.e.\/}, entropy-conserving (we assume baryon number conservation for $T<150$\,MeV). However, the inverse multi-particle decay process of hadrons decouples earlier than the two-to-one reactions we consider here. In this case, multiparticle decays add to entropy inventory while `mother' populations are kept in equilibrium by two-to-one reactions. The chemical potential change according to Eq.\,(\ref{muBeq}) if the present day value we used is different from that to be considered in the hadron era. To describe this novel situation kinetic theory for entropy production needs to be developed, a topic we hope to address in future work.

A related topic of future interest is another mechanism for the formation of (partial) strangeness chemical nonequilibrium: Reactions described above also change in the number of particles in detailed balance equation; the back reaction requiring additionally two particles becomes too slow. This was proposed as the origin of electron, positron decoupling in dense medium and source of chemical nonequilibrium in QED plasma~\cite{Aksenov:2007,Aksenov:2010}.

The primary conclusion of this first study of strangeness production and content in the early Universe, following on QGP hadronization, is that the relevant temperature domains indicate a complex interplay between baryon and meson (strange and non-strange) abundances and non-trivial decoupling from equilibrium for strange and non-strange mesons. We believe that this work contributes to the opening of a new and rich domain in the study of the Universe evolution in the future. This work is prerequisite for deeper understanding of heavier flavors, for bottom see~\cite{Yang:2020nne}; we plan to return to the case of charm flavor soon.\\[0.2cm] 

\textbf{Acknowledgment:} The research of CTY was in part supported by the US Department of Energy under Grant Contract DESC0012704 to the Brookhaven National Laboratory.
%%%%%%%%%%%%%%%%%%%%%%%%%%%%%%%%%%%%%%%%%%%%%%%%%%%%%%%%%%%%%%%%%%%

%%%%%%%%%%%%%%%%%%%%%%%%%%%%%%%%%%%%%%%%%%%%%%%%%%%%%%%%%%%%%%%%%%%

%%%%%%%%%%%%%%%%%%%%%%%%%%%%%%%%%%%%%%%%%%%%%%%%%%%%%%%%%%%%%%%%%%%

\end{document}